\documentclass[12pt]{iopart}
\usepackage[latin1]{inputenc}
\usepackage{amssymb}
\usepackage{graphicx}

\begin{document}

\title[Sublinear dispersive conductivity in PEI] {Sublinear dispersive conductivity in polyetherimides by the electric modulus formalism}

\author{M. Mudarra, J. Sellarès, J. C. Cañadas, J. A. Diego}
\address{Dept. Física i Enginyeria Nuclear, Universitat Polit{è}cnica de Catalunya,\\
Campus de Terrassa, C. Colom 11, Terrassa E-08222 (Spain)}
\ead{jordi.sellares@upc.es}

\begin{abstract}
It can be seen by Dynamic Electrical Analysis that the electrical properties of polyetherimide at temperatures above the glass transition are strongly influenced by space charge. We have studied space charge relaxation in two commercial grades of polyetherimide, Ultem 1000 and Ultem 5000, using this technique. The electric modulus formalism has been used to interpret their conductive properties. In both grades of polyetherimide, asymmetric Argand plots are observed, which are related to a sublinear power-law dependency ($\omega^{n}$ with $n<1$) in the real part of the conductivity. This behaviour is attributed to correlated ion hopping. The imaginary part of the electric modulus exhibits a peak in the low frequency range associated to conduction. Modelling of this peak allows us to obtain the dependence, among other parameters, of the conductivity ($\sigma_0$), the fractional exponent ($n$) and the crossover frequency ($\omega_\mathrm{p}$) on the temperature. The $\alpha$ relaxation, that appears at higher frequencies, has also to be modelled since it overlaps the conductivity relaxation.  
The study of the parameters in terms of the temperature allows us to identify the ones that are thermally activated. The difference between the conductivity relaxation time and the Maxwell relaxation time indicates the presence of deep traps.
The coupling model points out that the correlation of the ionic motion diminishes with temperature, probably due to increasing disorder due to thermal agitation.
\end{abstract}


\section{Introduction}

Polyetherimide (PEI) is an amorphous thermoplastic resin with a high
glass transition temperature that was developed by General Electric. 
Because of its advantageous mechanical, thermal,
and electrical properties, this resin is suitable for industrial applications,
such as microwave devices, high performance electrical devices and
biomedical applications. For many of these applications it is important 
to know the relaxational behaviour of the material to foresee its response. 
Among these relaxations, those related to space charge play an especially 
important role in the electrical response of the material. 

\begin{figure}[htbp]
\begin{center}
\includegraphics[clip,width=8.2cm]{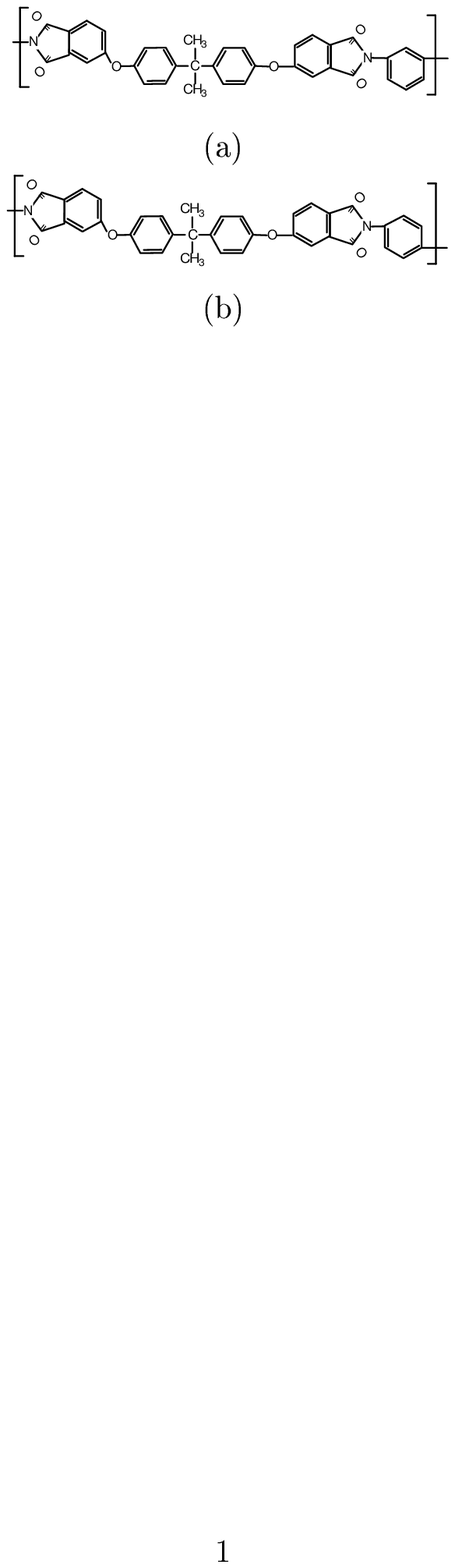}
\end{center}
\caption{\label{fig1} Chemical structures of Ultem 1000 (a) and Ultem 5000
(b)}
\end{figure}

Our study is focused on two commercial grades of PEI Ultem 1000 and
Ultem 5000. These resins have similar chemical structure, but Ultem
1000 is meta-linked whereas Ultem 5000 is para-linked (Fig.~\ref{fig1}).
Ultem 5000 has enhanced chemical stability versus acids
and organic solvents. 

Previous studies show that the dielectric strength of Ultem 5000 is higher
than in the case of Ultem 1000 over a wide temperature range and
the value of this magnitude diminishes with the temperature in
both grades \cite{ZEB98}. The differences are attributed to the remnant
electrical field due to space charge, as charge injection from the
electrode is more efficient in the case of Ultem 1000. Krause, Yang
and Sessler have observed that in corona charged samples surface charge
decays faster in the case of Ultem 5000 and that this grade is more
sensitive to the change of the polarity of the applied field \cite{KRA98}.
They attribute these differences to a more continuous band structure
due to the existence of an ordered morphology on Ultem 5000 films.

It is also known that the
loss factor of Ultem 1000 increases sharply for temperatures close
to the glass transition temperature due to conductive processes \cite{DIA98}. 
In the case of thermally stimulated
depolarization currents (TSDC) measurements, the relaxation of space
charge is associated with the $\rho$ relaxation, which appears
at temperatures higher than the $\alpha$ relaxation, associated to the glass transition
\cite{BEL98A}. This peak was studied by means of the
general order kinetic model \cite{CHEN,MUD97,MUD98,MUD99}, that provides
information about the relaxation mechanism for space charge and it
was concluded that recombination was the most likely relaxation mechanism
and that the depth of the traps is approximately 2.6~eV \cite{BEL98A}.

In this paper we present a relaxational study of the conductivity of
polyetherimide to improve the understanding of the mechanisms
related to space charge in this material. The $\alpha$ relaxation has also been 
studied since both relaxations are so close that they overlap and cannot be 
studied on their own. 

Data will be obtained by means of the dynamic electrical analysis (DEA) technique 
and interpreted using the electric modulus formalism \cite{DMOY98,DPIS97,DMAC02,hodge05}. 
The electric modulus ($M^\ast$) can be obtained from the permittivity ($\epsilon^\ast$) through
\begin{equation}
M^\ast = \left( \epsilon^\ast \right)^{-1}.
\end{equation}
The conductivity relaxation appears as a flex in the imaginary part of the permittivity. Instead, 
in the electric modulus it appears as peak allowing an easier interpretation of the data. The electric modulus formalism has been used to study the conductivity relaxation in polymers \cite{DMUD00,lu06}, glasses \cite{lanfredi02}, crystals \cite{rivera01}, ceramics \cite{liu03} and composites \cite{psarras03,migahed04}, among other materials. 

We expect to find a sublinear frequency
dispersive AC conductivity as in many other systems \cite{JONSCHER}. In this case the
real part of the conductivity $\sigma'(\omega)$ can be expressed
as
\begin{equation}
\sigma'(\omega)=\sigma_{0}+A\omega^{n}
\label{eq:realcond}
\end{equation}
where $\sigma_{0}$ is the DC conductivity, $A$ is a temperature
dependent parameter and $n$ is a fractional exponent which ranges
between 0 and 1 and has been interpreted by means of many body interactions
among charge carriers \cite{JONSCHER2}. 

This behaviour, termed universal dynamic response,
has been observed in highly disordered materials like ionically conducting
glasses, polymers, amorphous semiconductors and also in doped crystalline
solids \cite{lanfredi02,rivera01,migahed04,DSID95B,DSID97,LEO98,anantha05}. 
Equation~\ref{eq:realcond}
can be derived from the universal dielectric response function \cite{JONN80}
for the dielectric loss of materials with free hopping carriers. 

With regards to the $\alpha$ relaxation, we will consider that the relaxation time follows the Havriliak--Negami equation and has Vogel--Tammann--Fulcher dependence on temperature.

It will be of particular interest to find the dependence of the obtained parameters on the temperature \cite{DALM82}. All in all, the parameters that result from the modelling will allow us to characterize the relaxational behaviour of Ultem 1000 and Ultem 5000 above $T_\mathrm{g}$ and to obtain information about the space charge mechanisms in polyetherimide.

\section{Experimental}

Two grades of PEI, Ultem 1000 and Ultem 5000, were supplied by General
Electric. DSC measurements indicate that their glass transition temperatures
($T_{\mathrm{g}}$) are: $T_{\mathrm{g}} \approx 220$~$^\circ$C
in the case of Ultem 1000 and $T_{\mathrm{g}} \approx 235$~$^\circ$C
for ultem 5000. 

Samples of Ultem 1000 and Ultem 5000 were cut from
sheets of 125~$\mu$m in square portions with a side of 25~mm. The samples were placed between gold plated electrodes with radius 20~mm and measured using a Novocontrol BDS40 dielectric spectrometer with
a Novotherm temperature control system. 

The real and imaginary parts of the
electrical permittivity were measured at several frequencies between $10^{-2}$~Hz and $10^6$~Hz 
at isothermal steps of 5~$^\circ$C each at temperatures ranging between 230~$^\circ$C and 285~$^\circ$C in the case of Ultem 1000 and between 255~$^\circ$C and 290~$^\circ$C in the case of Ultem 5000. These temperature ranges are above the respective glass transition of the PEI
grades in order to get a substantial contribution of the conductive
processes.

\section{Results and discussion}

\begin{figure}
\begin{center}
\includegraphics[clip,width=8.6cm]{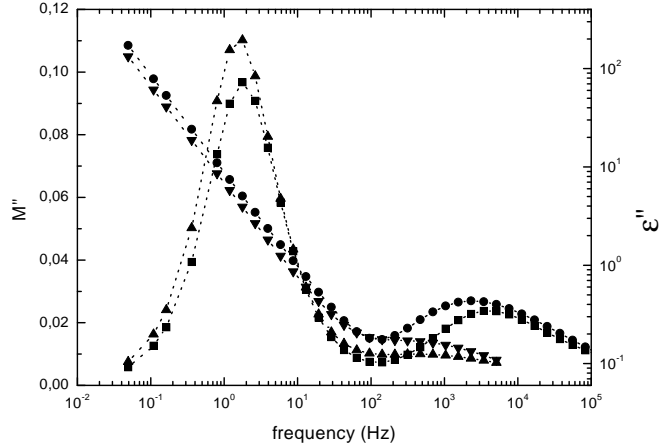} 
\end{center}
\caption{\label{fig2}Imaginary part of the electric modulus at $T=255\;^{o}\mathrm{C}$
of Ultem 1000 ($\blacksquare$) and Ultem 5000 ($\blacktriangle$).
Loss factor of Ultem 1000 ($\bullet$) and Ultem 5000 ($\blacktriangledown$)
at the same temperature.}
\end{figure}

The loss factor ($\varepsilon''$) and the imaginary part of the electric
modulus ($M''$) of Ultem 1000 and Ultem 5000 are plotted in figure~\ref{fig2}
as a function of the frequency for a temperature $T=255\;^{\mathrm{o}}\mathrm{C}$,
which lies above $T_{\mathrm{g}}$ in both cases. It can be seen that the conductive
processes in both materials result in a sharp increase of the loss
factor at low frequencies. In the case of the imaginary part of the
electric modulus, these effects are evidenced by a peak. This is certainly easier to model \cite{hodge05,AVAK92}. For this reason we have choosed the electric modulus formalism to study the conduction relaxation.

\begin{figure}
\begin{center}
\includegraphics[clip,width=8.6cm]{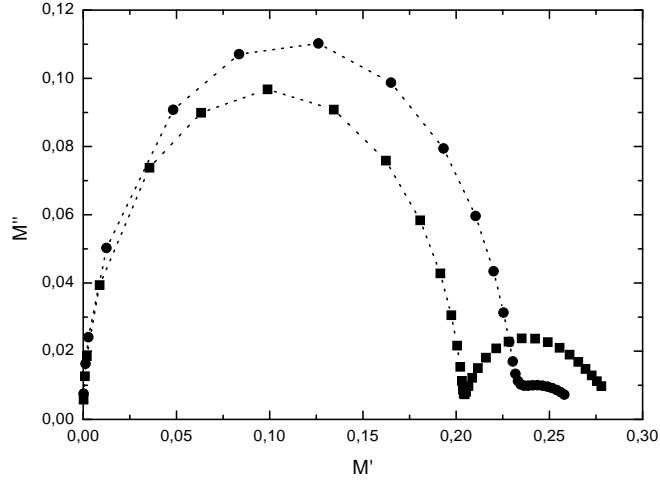} 
\end{center}
\caption{\label{fig3} Argand's plot of the electric modulus $M^{\ast}$ of
Ultem 1000 ($\blacksquare$) and Ultem 5000 ($\bullet$) at $T=255\;^{o}\mathrm{C}$.
(Frequency of plotted data increases from left to right}
\end{figure}

In figure~\ref{fig3}
we have plotted the electric modulus in Argand's plane, for temperatures
close to the glass transition temperature of Ultem 1000. The corresponding
plots of the electric modulus show two arcs, i.e., the conductive
processes result in an arc for low frequencies and for higher frequencies
we can observe another arc that can be associated with the $\alpha$ relaxation.
Depressed arcs are observed in the case of the relaxation associated with
conductive processes that can be related to a dispersive regime \cite{DMAC}.  

\begin{figure}
\begin{center}
\includegraphics[clip,width=8.6cm]{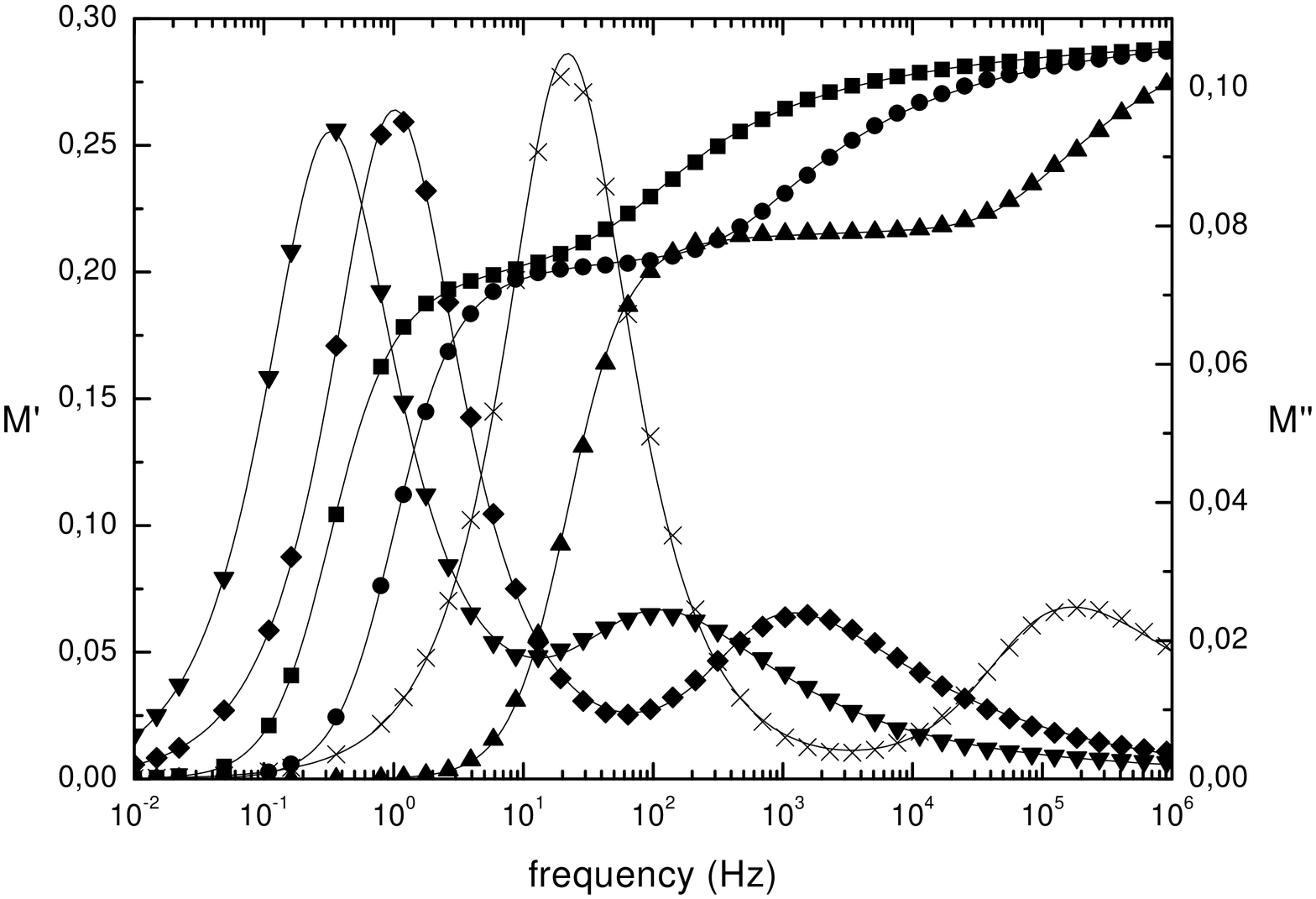}
\end{center} 
\caption{\label{fig4} Real and Imaginary parts of the electric modulus of
Ultem 1000 versus the frequency for several temperatures. Real part:
$\blacksquare$, $240\;^{o}\mathrm{C}$; $\bullet$, $250\;^{o}\mathrm{C}$;
$\blacktriangle$, $280\;^{o}\mathrm{C}$; Imaginary part: $\blacktriangledown$,
$240\;^{o}\mathrm{C}$; $\blacklozenge$, $250\;^{o}\mathrm{C}$;
$\times$, $280\;^{o}\mathrm{C}$. The symbols are measured values
and the continuous curves are the calculated values using the values
of table~\ref{table1} resulting from the fitting process to equation~\ref{eq:modulus}}
\end{figure}

In figure 
\ref{fig4}
we have plotted the peak associated with the conductive
process in Ultem 1000 for several temperatures above the glass transition.
It can be noted that the peak shifts to higher frequencies with the
temperature.

To study the charge transport process at these temperatures we have
assumed a sublinear frequency dispersive AC conductivity, as depressed
arcs observed in figure~\ref{fig3} for low frequencies can be associated
with a dispersive regime. Power--law dependencies of conductivity,
as in the case of equation \ref{eq:realcond}, imply a power--law
dependence of the form $(\mathrm{j}\;\omega)^{n}$ for the complex
conductivity\cite{LEO97}. Therefore, this magnitude can be written:
\begin{equation}
\sigma^{\ast}(\omega)=\sigma_{0}+A(\mathrm{j}\;\omega)^{n}+\mathrm{j}\;\omega\varepsilon_{0}\varepsilon_{\infty \mathrm{C}}\label{eq:complcond1}
\end{equation}
 where $\varepsilon_{\infty \mathrm{C}}$ refers to the permittivity at high
frequency. A crossover frequency $\omega_{\mathrm{p}}$ can be defined as 
\begin{equation}
\omega_{\mathrm{p}}^{n} \equiv \sigma_{0}/A
\label{adete}
\end{equation}
so that equation \ref{eq:complcond1} can be rewritten as 
\begin{equation}
\sigma^{\ast}(\omega)=\sigma_{0}+\sigma_{0}\left(\mathrm{j}\frac{\omega}{\omega_{\mathrm{p}}}\right)^{n}+\mathrm{j}\;\omega\varepsilon_{0}\varepsilon
_{\infty \mathrm{C}}\label{eq:complcond}
\end{equation}

This frequency $\omega_{\mathrm{p}}$ is associated with the crossover from
the power-law dependence observed at high frequency to a frequency
independent DC regime that occurs at low frequencies. Finally the
contribution to the permittivity is 
\begin{equation}
\varepsilon_{\mathrm{C}}^{\ast}=-\frac{j\sigma^{\ast}}{\varepsilon_{0}\omega}\label{eq:cond}
\end{equation}

The $\alpha$ relaxation can be modelized by means of Havriliak-Negami
equation 
\begin{equation}
\varepsilon_{\mathrm{HN}}^{\ast}=\varepsilon_{\infty \mathrm{HN}}+\frac{\Delta\varepsilon}{\left(1+(j\omega\tau_{\mathrm{HN}})^{\alpha_{\mathrm{HN}}}\right)^{\beta_{\mathrm{HN}}}}\label{eq:hn}
\end{equation}
 where $\varepsilon_{\infty \mathrm{HN}}$ refers to the permittivity at high
frequencies and 
\begin{equation}
\Delta\varepsilon=\varepsilon_{\infty \mathrm{C}}-\varepsilon_{\infty \mathrm{HN}}
\label{strength}
\end{equation}
is the relaxation strength. Therefore the electric modulus over the
frequency range considered can be expressed as: 
\begin{equation}
M^{\ast}=\left(\varepsilon_{\mathrm{C}}^{\ast}+\varepsilon_{\mathrm{HN}}^{\ast}\right)^{-1}\label{eq:modulus}
\end{equation}

Electric modulus versus frequency data have been fitted to equation
\ref{eq:modulus}. The real and imaginary parts of the electric modulus,
$M^{\ast}(\omega)$, were calculated from the complex permittivity
and were fitted to the real and imaginary parts of the electric modulus
given by equation~\ref{eq:modulus} simultaneously. Eight independent
parameters were used in the fitting process: $\sigma_{0}$, $\omega_{\mathrm{p}}$,
$\varepsilon_{\infty \mathrm{C}}$, $n$, $\varepsilon_{\infty \mathrm{HN}}$ , $\tau_{\mathrm{HN}}$,
$\alpha_{\mathrm{HN}}$ and $\beta_{\mathrm{HN}}$. 

\begin{figure}
\begin{center}
\includegraphics[clip,width=8.6cm]{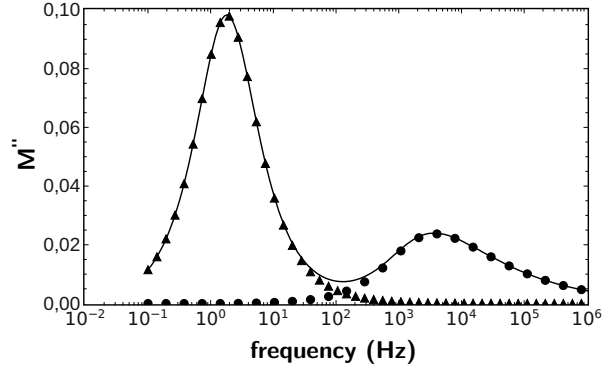} 
\end{center}
\caption{\label{fig5}Imaginary part of the electric modulus at $T=255\;^{o}\mathrm{C}$
of Ultem 1000: line is the value calculated by means of equation~\ref{eq:modulus};
symbols indicate the separate contributions of conductive process
given by equation~\ref{eq:cond} ($\blacktriangle$) and main relaxation
$\alpha$ ($\bullet$) given by equation~\ref{eq:hn}. Parameters
$\sigma_{0}$, $\omega_{\mathrm{p}}$, $\varepsilon_{\infty \mathrm{C}}$,
$n$ characterize conductive process, whereas that parameters $\varepsilon
_{\infty \mathrm{HN}}$
, $\tau_{\mathrm{HN}}$, $\alpha_{\mathrm{HN}}$ and $\beta_{\mathrm{HN}}$ characterize main
relaxation $\alpha$}
\end{figure}

The study of the two contributions
to electric modulus, given by equation~\ref{eq:cond} and equation~\ref{eq:hn},
indicates that the first contribution
(which is characterized by four of the parameters: $\sigma_{0}$,
$\omega_{\mathrm{p}}$, $\varepsilon_{\infty \mathrm{C}}$, $n$) helps to
explain appropriately the contribution of conductive processes, whereas
that the second contribution (which is characterized by the four remaining
parameters, $\varepsilon_{\infty \mathrm{HN}}$ , $\tau_{\mathrm{HN}}$, $\alpha_{\mathrm{HN}}$
and $\beta_{\mathrm{HN}}$) explains main relaxation $\alpha$. 
Both contributions are plotted separately in figure~\ref{fig5}.
It can be noted that each process (and, therefore, each set of four parameters) mainly
contributes in a different frequency range, and the whole set of eight
parameters is required to explain adequately the experimental response
observed over the whole frequency range.

In this work we have used simulated annealing to carry out the fitting
process. This method has been successfully used in the analysis of
thermally stimulated depolarization currents~\cite{DLAR99,DGRI97}
and dielectric spectroscopy data~\cite{DBEL99}. The values of the
parameters obtained are shown in tables \ref{table1}, \ref{table2},
\ref{table3} and \ref{table4}. A good agreement between experimental
and calculated data (symbols and continuous line respectively) has
been obtained as it can be seen in figure \ref{fig4}.

\begin{table}
\caption{\label{table1} Parameters associated with conductive processes: Ultem
1000 case. All parameters are fit results.}
\begin{center}
\begin{tabular}{cccccc}
\hline 
T ($^{o}\mathrm{C}$)  & $\sigma_{0}$ ($\Omega^{-1}\mathrm{cm}^{-1}$)  & $\omega_{\mathrm{p}}$ (s$^{-1}$)  & $n$  & $\varepsilon_{\infty \mathrm{C}}$ \tabularnewline
\hline 
230  & 0.271$\times10^{-10}$  & 14.9  & 0.678  & 5.00 \tabularnewline
235  & 0.490$\times10^{-10}$  & 59.8  & 0.596  & 5.00 \tabularnewline
240  & 0.896$\times10^{-10}$  & 90.8  & 0.628  & 4.91 \tabularnewline
245  & 0.156$\times10^{-9}$  & 375.5  & 0.546  & 4.92 \tabularnewline
250  & 0.272$\times10^{-9}$  & 787.4  & 0.524  & 4.88 \tabularnewline
255  & 0.473$\times10^{-9}$  & 0.590$\times10^{4}$  & 0.429  & 4.88 \tabularnewline
260  & 0.803$\times10^{-9}$  & 0.165$\times10^{5}$  & 0.407  & 4.85 \tabularnewline
265  & 0.133$\times10^{-8}$  & 0.567$\times10^{5}$  & 0.344  & 4.81 \tabularnewline
270  & 0.201$\times10^{-8}$  & 0.552$\times10^{6}$  & 0.202  & 4.76 \tabularnewline
275  & 0.345$\times10^{-8}$  & 0.430$\times10^{6}$  & 0.315  & 4.72 \tabularnewline
280  & 0.483$\times10^{-8}$  & 0.588$\times10^{7}$  & 0.144  & 4.64 \tabularnewline
285  & 0.748$\times10^{-8}$  & 0.726$\times10^{6}$  & 0.217  & 4.00 \tabularnewline
\hline 
\end{tabular}
\end{center}
\end{table}

\begin{table}
\caption{\label{table2} Parameters associated with conductive processes: Ultem
5000 case. All parameters are fit results.}
\begin{center}
\begin{tabular}{cccccc}
\hline 
T ($^{o}\mathrm{C}$)  & $\sigma_{0}$ ($\Omega^{-1}\mathrm{cm}^{-1}$)  & $\omega_{\mathrm{p}}$ (s$^{-1}$)  & $n$  & $\varepsilon_{\infty \mathrm{C}}$ \tabularnewline
\hline 
255  & 0.368$\times10^{-9}$  & 0.181$\times10^{4}$  & 0.526  & 4.28 \tabularnewline
260  & 0.577$\times10^{-9}$  & 0.204$\times10^{4}$  & 0.580  & 4.25 \tabularnewline
265  & 0.904$\times10^{-9}$  & 0.824$\times10^{4}$  & 0.507  & 4.24 \tabularnewline
270  & 0.139$\times10^{-8}$  & 0.569$\times10^{5}$  & 0.373  & 4.22 \tabularnewline
275  & 0.213$\times10^{-8}$  & 0.144$\times10^{6}$  & 0.365  & 4.17 \tabularnewline
280  & 0.322$\times10^{-8}$  & 0.346$\times10^{6}$  & 0.332  & 4.10 \tabularnewline
285  & 0.499$\times10^{-8}$  & 0.198$\times10^{6}$  & 0.381  & 3.99 \tabularnewline
290  & 0.624$\times10^{-8}$  & 0.120$\times10^{6}$  & 0.303  & 3.72 \tabularnewline
\hline 
\end{tabular}
\end{center}
\end{table}

\begin{table}
\caption{\label{table3} Parameters associated with $\alpha$ relaxation: Ultem
1000 case. All parameters are fit results except $\Delta\varepsilon$ that is 
calculated through equation~\ref{strength}.}
\begin{center}
\begin{tabular}{cccccc}
\hline 
T ($^{o}\mathrm{C}$)  & $\varepsilon_{\infty \mathrm{HN}}$  & $\tau_{\mathrm{HN}}$ (s)  & $\alpha_{\mathrm{HN}}$  & $\beta_{\mathrm{HN}}$  & $\Delta\varepsilon$\tabularnewline
\hline 
230  & 3.51  & 0.119  & 0.655  & 0.748  & 1.49\tabularnewline
235  & 3.50  & 0.189$\times10^{-1}$  & 0.729  & 0.627  & 1.50\tabularnewline
240  & 3.48  & 0.393$\times10^{-2}$  & 0.813  & 0.549  & 1.43\tabularnewline
245  & 3.46  & 0.123$\times10^{-2}$  & 0.865  & 0.494  & 1.46\tabularnewline
250  & 3.45  & 0.377$\times10^{-3}$  & 0.907  & 0.462  & 1.43\tabularnewline
255  & 3.43  & 0.138$\times10^{-3}$  & 0.899  & 0.464  & 1.45\tabularnewline
260  & 3.40  & 0.575$\times10^{-4}$  & 0.922  & 0.443  & 1.45\tabularnewline
265  & 3.37  & 0.245$\times10^{-4}$  & 0.929  & 0.436  & 1.44\tabularnewline
270  & 3.35  & 0.111$\times10^{-4}$  & 0.926  & 0.439  & 1.41\tabularnewline
275  & 3.32  & 0.564$\times10^{-5}$  & 0.925  & 0.443  & 1.40\tabularnewline
280  & 3.25  & 0.291$\times10^{-5}$  & 0.921  & 0.435  & 1.39\tabularnewline
285  & 3.19  & 0.163$\times10^{-5}$  & 0.929  & 0.449  & 1.30\tabularnewline
\hline 
\end{tabular}
\end{center}
\end{table}

\begin{table}
\caption{\label{table4} Parameters associated with $\alpha$ relaxation: Ultem
5000 case. All parameters are fit results except $\Delta\varepsilon$ that is 
calculated through equation~\ref{strength}.}
\begin{center}
\begin{tabular}{cccccc}
\hline 
T ($^{o}\mathrm{C}$)  & $\varepsilon_{\infty \mathrm{HN}}$  & $\tau_{\mathrm{HN}}$ (s)  & $\alpha_{\mathrm{HN}}$  & $\beta_{\mathrm{HN}}$  & $\Delta\varepsilon$\tabularnewline
\hline 
255  & 3.65  & 0.781$\times10^{-3}$  & 0.803  & 0.365  & 0.63\tabularnewline
260  & 3.61  & 0.248$\times10^{-3}$  & 0.864  & 0.309  & 0.64\tabularnewline
265  & 3.59  & 0.840$\times10^{-4}$  & 0.845  & 0.315  & 0.65\tabularnewline
270  & 3.54  & 0.367$\times10^{-4}$  & 0.854  & 0.284  & 0.68\tabularnewline
275  & 3.50  & 0.150$\times10^{-4}$  & 0.835  & 0.298  & 0.67\tabularnewline
280  & 3.43  & 0.689$\times10^{-5}$  & 0.822  & 0.299  & 0.67\tabularnewline
285  & 3.33  & 0.356$\times10^{-5}$  & 0.839  & 0.284  & 0.66\tabularnewline
290  & 3.11  & 0.208$\times10^{-5}$  & 0.847  & 0.283  & 0.61\tabularnewline
\hline 
\end{tabular}
\end{center}
\end{table}

\begin{figure}
\begin{center}
\includegraphics[clip,width=8.2cm]{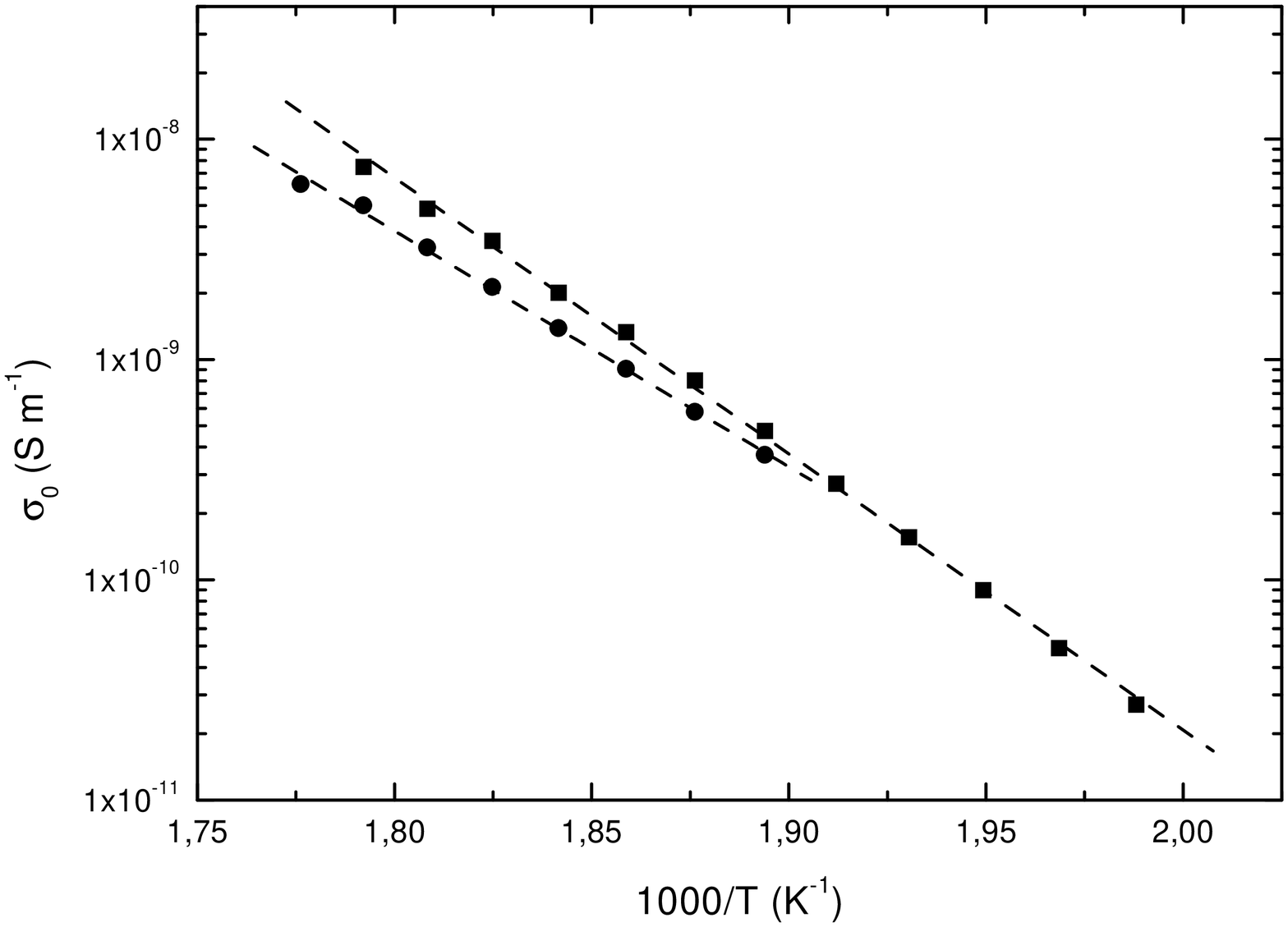}
\end{center} 
\caption{\label{fig6} Arrhenius plot of DC conductivity $\blacksquare$, Ultem
1000; $\bullet$, Ultem 5000}
\end{figure}

The DC conductivity ($\sigma_0$) of both grades of polyetherimide increases with
the temperature as it can be seen in tables~\ref{table1} and ~\ref{table2}. Figure~\ref{fig6}
shows an Arrhenius plot of the conductivity where it can be seen that it is thermally activated.
This increase has been attributed to an increase of carrier mobility \cite{DMUD00}.
It can be fitted to 
\begin{equation}
\sigma_{0}=\sigma_{0\mathrm{PF}}\exp(E_{\mathrm{a}}/kT)
\label{eq:thermallyactivated}
\end{equation}
The activation energies and preexponential factors can be seen in
table \ref{table5}. Ultem 1000 presents a slightly higher activation
energy for the DC conductivity.

Concerning the relaxation time associated with space charge relaxation,
it has been proposed that the crossover frequency can be associated
with a characteristic time $\tau_{\mathrm{p}}$ by means of $\tau_{\mathrm{p}}=2\pi/\omega_{\mathrm{p}}$.
Experimental evidence has been given to support the idea that
this characteristic time is actually the same time that an average
relaxation time $\langle\tau\rangle$\cite{LEO97}, which can be defined
in terms of the area under the Kohlrausch-Williams-Watts (KWW) function
as 
\begin{equation}
\langle\tau\rangle=\int_{0}^{\infty}\Phi(t)\; dt=\frac{\Gamma(1/\beta)\tau^{\ast}}{\beta}
\end{equation}
 where $\tau^{\ast}$ and $\beta$ are the parameters that define the KWW response function 
\begin{equation}
\Phi(t)=\exp[-(t/\tau^{\ast})^{\beta}]\label{eq:kww}
\end{equation}

This average relaxation time $\langle\tau\rangle$ corresponds to the
Maxwell relaxation time, this is, the time that an out--of--equilibrium conductor needs to reach electrostatic equilibrium.
It is related to the DC conductivity according to the expression \cite{hodge05} 
\begin{equation}
\langle\tau\rangle=\frac{\varepsilon_{0}\varepsilon_{\infty}}{\sigma_{0}}
\label{eq:art}
\end{equation}

\begin{figure}
\begin{center}
\includegraphics[clip,width=8.6cm]{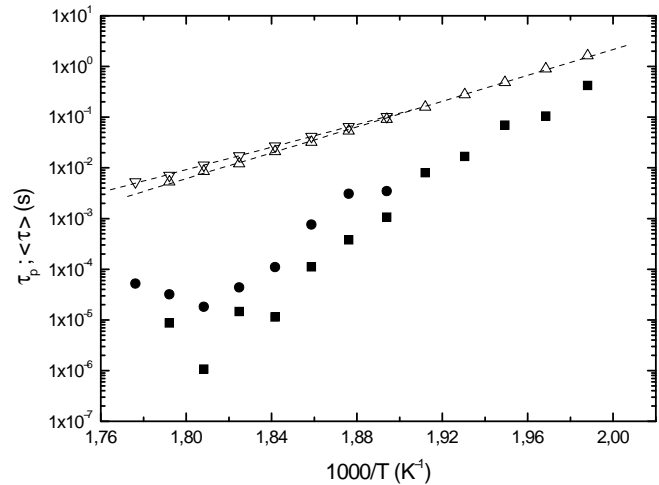} 
\end{center}
\caption{\label{fig7} Characteristic time $\tau_{\mathrm{p}}$ (associated with crossover
frequency $\tau_{\mathrm{p}}=\frac{2\pi}{\omega_{\mathrm{p}}})$ :$\blacksquare$ ,
Ultem 1000; $\bullet$, Ultem 5000. Maxwell relaxation time $\langle\tau\rangle$:
$\triangle$ Ultem 1000; $\bigtriangledown$ Ultem 5000}
\end{figure}

We have calculated the values of $\langle\tau\rangle$ and $\tau_{\mathrm{p}}$. An attempt to compare both magnitudes in an Arrhenius plot can be seen in figure~\ref{fig7}. 
As it can be seen, both quantities represent a thermally activated relaxation
time of the form 
\begin{equation}
\tau=\tau_{\mathrm{PF}}\exp(E_{\mathrm{a}}/kT)
\label{eq:thermallyactivated2}
\end{equation}
The activation energies and preexponential factors can be seen in
table \ref{table5}.
\begin{table}
\caption{Preexponential factors ($PF$) and activation energies ($E_{\mathrm{a}}$)
of conductivity ($\sigma_0$) and relaxation times ($\langle\tau\rangle$
and $\tau_{\mathrm{p}}$), obtained from the corresponding Arrhenius plots\label{table5}}
\begin{center}
\begin{tabular}{crr}
\hline 
 & Ultem 1000  & \tabularnewline
Magnitude  & $\log_{10}(\mathrm{PF}/1\mathrm{s})$  & $E_{\mathrm{a}}$ (eV) \tabularnewline
\hline 
$\sigma_0$  & $14.4$  & $1.08$ \tabularnewline
$\langle\tau\rangle$  & $-25.1$  & $1.10$ \tabularnewline
$\tau_{\mathrm{p}}$  & $-56.6$~s  & $2.44$\tabularnewline
\hline 
 & Ultem 5000  & \tabularnewline
Magnitude  & $\log_{10}(\mathrm{PF}/1\mathrm{s})$  & $E_{\mathrm{a}}$ (eV) \tabularnewline
\hline 
$\sigma_0$  & $10.8$  & $0.92$ \tabularnewline
$\langle\tau\rangle$  & $-22.1$  & $1.96$ \tabularnewline
$\tau_{\mathrm{p}}$  & $-58.4$  & $2.56$ \tabularnewline
\hline 
\end{tabular}
\end{center}
\end{table}
For both types of relaxation time, the activation energy in the case of Ultem~5000 is higher. It can also be noticed that the activation energy of $\tau_p$ is higher than the one of $\langle \tau \rangle$.

It should be clear that these quantities do not represent the same magnitude since their values differ in several orders
of magnitude, specially at high temperatures. The agreement between
these values was used by León \textit{et al.} to support the hypothesis
of a common origin for both DC and AC regimes \cite{LEO97}. Nevertheless, in the following lines we will explain how our results
can be reconciled with the idea that both quantities are, at least, related.

In our opinion, the physical meaning of $\tau_{\mathrm{p}}$ is just an estimation of
the time that separates the relaxation processes for which predominates
DC conduction ($\tau>\tau_{\mathrm{p}}$) or predominates AC ($\tau<\tau_{\mathrm{p}}$)
conduction. 

The mechanism that leads to DC regime are long range displacements \cite{DSID95B} where the carriers move in the same direction
but at higher frequencies short range hopping of the ion occurs, which is viewed as
a correlated motion in which the ion performs several reiterated forward--backward
hops before completing any successful forward displacement. Reiterative
hopping is the origin of the dispersive regime and it occurs when the frequency
is higher than the crossover frequency below which successful hops can be completed.

The high value of $\langle\tau\rangle$ that we have
found at high temperatures, in the case of macromolecular materials
can be explained on the basis that the microscopic processes that
lead to ionic conduction may not be reduced to mere hops over a barrier
between adjacent sites. Amorphous polymers, such as PEI, are disordered
materials and charge may be located in deep traps. This kind of charge 
plays a relevant role in microscopic relaxation processes. When
the temperature is increased deeply trapped charge may get not enough
thermal energy to jump over the potential barrier, but to reach a
intermediate energy level, resulting in a more complex relaxation behaviour.

If one assumes that the only process that contributes to electrical
conduction is hopping, the crossover frequency can be associated with
the frequency below which ions can follow the variations of the applied
field, so that their characteristic relaxation time is shorter than
the applied field period. At such frequencies, DC conductivity determines 
the relaxation process and, therefore, the Maxwell time determines the 
crossover frequency. 

But if deep traps are present, ions deeply trapped 
have a longer relaxation time, so that their contribution
should be evident at frequencies below the crossover frequency and
their effect should represent a slowing down of space charge relaxation
process. On the other hand, ions located in deep traps can not follow
the field oscillations at higher frequencies and they do not contribute
to the relaxation process. Therefore, the presence of deep traps can
be associated with a slowing down of the relaxation process, that
results in an apparent Maxwell time longer than the relaxation time
that corresponds to the crossover frequency. The higher value of $\langle\tau\rangle$
and the higher activation energy of $\tau_{\mathrm{p}}$ can thus be explained by
the presence of deep traps.

\begin{figure}
\begin{center}
\includegraphics[clip,width=8.6cm]{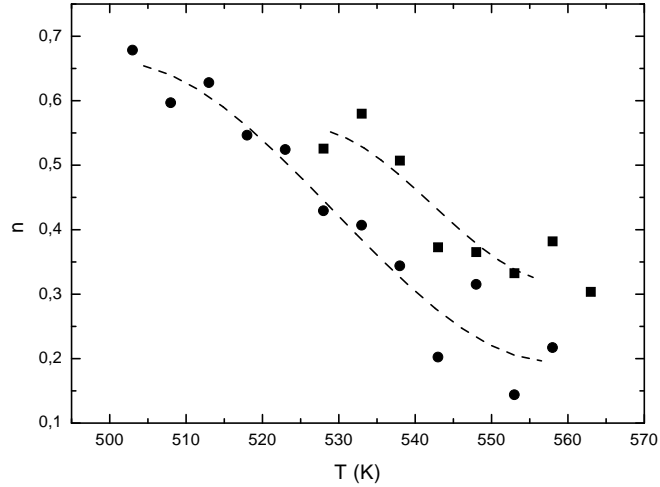} 
\end{center}
\caption{\label{fig8}Fractional exponent n as a function of temperature (Dashed
lines are a guide for the eye):$\blacksquare$ , Ultem 1000; $\bullet$,
Ultem 5000.}
\end{figure}

The temperature dependence of the fractional exponent $n$ is shown in figure~\ref{fig8}.
This parameter characterizes the power--law conduction regime, which
is associated with the slowing down of the relaxation process in the
frequency domain as a result of the cooperative effects, in the same
way as the KWW function does in the time domain. An important connection
between these two approaches stems from the coupling model of Ngai
and Kannert~\cite{DSID95}. This model predicts a power--law conductivity
associated with the KWW relaxation function (equation~\ref{eq:kww})
given by 
\begin{equation}
\sigma_{\mathrm{KWW}}=B\exp(-E_{\mathrm{a}}/kT)\;\omega^{1-\beta}
\end{equation}
Therefore, if any other contribution is sufficiently smaller than
that of $\sigma_{\mathrm{KWW}}$, then the conductivity of the material may
be described by $\sigma_{\mathrm{KWW}}(\omega)$ \cite{DSID95}. In that case, the sublinear
dispersive AC conductivity observed in polyetherimide can be associated
with a KWW relaxation mechanism with $\beta=1-n$ where $n$ is the
power--law exponent determined from $\sigma(\omega)$.

The stretched exponential relaxation
time has been associated with a slowing of the relaxation process
that results from correlated hopping. 
according to this interpretation, the stretched exponential 
parameter $\beta$ represents a correlation index of carrier motion.

One would expect $\beta$
to be close to zero for strongly correlated systems and close to 1
for random Debye--like hops. As we can see in figure~\ref{fig8}
the power--law exponent $n=1-\beta$ decreases with temperature for
both grades, indicating that correlation of ionic motion decreases
with temperature. This can be associated with the increasing disorder
due to thermal agitation of polymer chain segments.

In tables \ref{table3}
and \ref{table4} we can see the evolution of the fitting parameters
for the $\alpha$ relaxation. 
The $\alpha_{\mathrm{HN}}$ parameter indicates the non--exponentiality of the relaxation. This is, $\alpha_{\mathrm{HN}} = 1$ is a Debye--like relaxation and $\alpha_{\mathrm{HN}} < 1$ indicates that the relaxation extends over a wider range than a Debye relaxation. The non--exponentiality maybe due to a distribution of relaxation times or to cooperativity.
As the temperature increases and attains
values progressively further from the $T_{\mathrm{g}}$ of the material, the
value of $\alpha_{\mathrm{HN}}$ is closer to 1. In this case this is due to less cooperativity as the glass transition is left further away.
On the other hand, the parameter $\beta_{\mathrm{HN}}$ decreases, which means
that a broader range of relaxation modes are excited at temperatures
further from $T_{\mathrm{g}}$, but at the expense of a lower overall dielectrics
strength, as it can be seen in the tables. This is analogous to what happens in TSDC when a poling temperature far from the optimal one is employed.

\begin{figure}
\begin{center}
\includegraphics[clip,width=8.6cm]{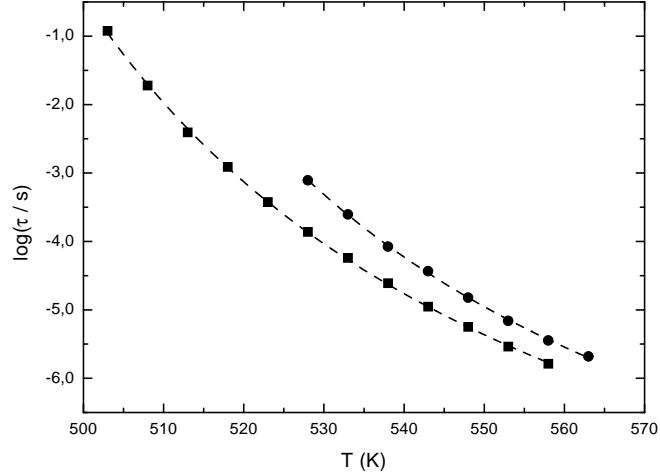} 
\end{center}
\caption{\label{fig9} Relaxation time $\tau_{\mathrm{HN}}$ as a function of temperature:
$\blacksquare$ , Ultem 1000; $\bullet$, Ultem 5000. Dashed lines
are curves fitted to VFT equation.}
\end{figure}

In the case of the $\alpha$ relaxation, figure \ref{fig9} 
shows that the relaxation time $\tau_{\mathrm{HN}}$ decreases with temperature following the
Vogel--Fulcher--Tammann (VFT) equation 
\begin{equation}
\tau_{\mathrm{HN}}=\tau_{\mathrm{HN}0}\frac{E_{\mathrm{W}}}{k_{\mathrm{B}}(T-T_{0})}\label{eq:VFT}
\end{equation}
for both PEI grades. The result of the fits can be seen in table~\ref{table6}.
\begin{table}
\caption{Preexponential factors $\tau_{\mathrm{HN}0}$, $E_{\mathrm{W}}$ and $T_{0}$ obtained
by fitting the relaxation time $\tau_{\mathrm{HN}}$ to VTF equation }
\label{table6} %
\begin{center}
\begin{tabular}{crrr}
\hline 
Grade  & $\tau_{\mathrm{HN}0}$ (s)  & $E_{\mathrm{W}}$ (eV)  & $T_{0}$ (K) \tabularnewline
\hline 
ULTEM 1000  & $1.43\times10^{-11}$  & 0.11  & 446 \tabularnewline
ULTEM 5000  & $4.95\times10^{-12}$  & 0.12  & 452 \tabularnewline
\hline 
\end{tabular}
\end{center}
\end{table}
This behaviour is typical of cooperative non-exponential
relaxations. This is not surprising since the $\alpha$ relaxation
is related to the glass transition of the material. The value of $T_0$ is lower in Ultem~1000 than in Ultem~5000. This is logical taking into account that they should follow a pattern similar to $T_{\mathrm{g}}$.

\section{Conclusions}

Conductive processes in two grades of commercial polyetherimide have
been studied and it has been found that they condition the electrical behaviour of
these materials at high temperatures and low frequencies. The electric
modulus formalism has been useful in order to interpret dynamic
electrical analysis data to characterize these processes.

The $\alpha$ relaxation has also been studied due to its proximity to the 
conductive relaxation that made impossible to study just the conduction on their own.

The dispersive conductivity observed in both grades of polyetherimides
can be explained by means of a sublinear frequency dispersive AC conductivity.
This behaviour is the result of correlated hopping. 

The DC conductivity is thermally activated, probably due to an increase of carriers.

Among the parameters studied, there is the crossover relaxation time $\tau_{\mathrm{p}}$. 
The physical meaning of $\tau_{\mathrm{p}}$ is just an estimation of
the time that separates the relaxation processes for which predominates
DC conduction ($\tau>\tau_{\mathrm{p}}$) or predominates AC ($\tau<\tau_{\mathrm{p}}$)
conduction. 

In the literature \cite{LEO97}, experimental evidence of the equivalence of the crossover relaxation time with the Maxwell relaxation time has been presented but in our case the presence of deep traps can produce a slowing down of the relaxation process. This results in an apparent Maxwell time longer than the crossover relaxation time.

Both relaxation times are thermally activated. This is to be expected in the case of the Maxwell relaxation time, since it is calculated from the DC conductivity. In the case of the crossover relaxation time, this fact reinforces the idea that it is physically related to the Maxwell relaxation time. The higher activation energy in the case of the crossover relaxation time can also be attributed to the presence of deep traps.

The power--law exponent $n$ decreases with temperature for
both grades, indicating that correlation of ionic motion decreases
with temperature. This can be associated with the increasing disorder
due to thermal agitation of polymer chain segments.

The behaviour of the $\alpha$ relaxation is typical of cooperative non-exponential
relaxations. This is not surprising since the $\alpha$ relaxation
is related to the glass transition of the material. As the temperature attains values further to $T_{\mathrm{g}}$ the response from the $\alpha$ relaxation becomes less cooperative even though a broader range of the relaxation is involved. 

\section*{Acknowledgements} This work has been partially supported by project 2009 SGR 01168 (AGAUR).


\section*{References}

\begin{thebibliography}{10}

\bibitem{ZEB98}
N.~Zebouchi, V.~H. Truong, R.~Essolbi, M.~Se-Ondoua, D.~Malec, N.~Vella,
  S.~Malrieu, A.~Toureille, F.~Schu{\'e}, and R.~G. Jones.
\newblock The electric breakdown behaviour of polyetherimide films.
\newblock {\em Polym. Int.}, 46(1):54--58, 1998.

\bibitem{KRA98}
E.~Krause, G.~M. Yang, and G.~M. Sessler.
\newblock Charge dynamics and morphology of {U}ltem 1000 and {U}ltem 5000 {PEI}
  grade films.
\newblock {\em Polym. Int.}, 46(1):59--64, 1998.

\bibitem{DIA98}
R.~D{\'{\i}}az-Calleja, S.~Friederichs, C.~Ja{\"{\i}}mes, M.~J. Sanchis,
  J.~Belana, J.~C. Ca{\~n}adas, J.~A. Diego, and M.~Mudarra.
\newblock Comparative study of mechanical and electrical relaxations in
  poly(etherimide). part 2.
\newblock {\em Polym. Int.}, 46(1):20--28, 1998.

\bibitem{BEL98A}
J.~Belana, J.~C. Ca{\~{n}}adas, J.~A. Diego, M.~Mudarra, R.~D\'{\i}az-Calleja,
  S.~Friedericks, C.~Jaïmes, and M.~J. Sanchis.
\newblock Comparative study of mechanical and electrical relaxations in
  poly(etherimide). part~1.
\newblock {\em Polym. Int.}, 46(1):11--19, 1998.

\bibitem{CHEN}
R.~Chen and Y.~Kirsh.
\newblock {\em Analisys of thermally stimulated processes}, chapter~2, pages
  35--92.
\newblock Pergamon Press, Oxford, 1981.

\bibitem{MUD97}
M.~Mudarra and J.~Belana.
\newblock Study of poly(methyl methacrylate) space charge relaxation by {TSDC}.
\newblock {\em Polymer}, 38:5815--5821, 1997.

\bibitem{MUD98}
M.~Mudarra, J.~Belana, J.~C. Ca{\~n}adas, and J.~A. Diego.
\newblock Polarization time effect on {PMMA} space charge relaxation by {TSDC}.
\newblock {\em J. Polym. Sci. Pol. Phys.}, pages 1971--1980, 1998.

\bibitem{MUD99}
M.~Mudarra, J.~Belana, J.~C. Ca{\~n}adas, and J.~A. Diego.
\newblock Windowing polarization: considerations to study the space charge
  relaxation in poly(methyl methacrylate) by thermally stimulated
  depolarization currents.
\newblock {\em Polymer}, 40:2659--2665, 1999.

\bibitem{DMOY98}
C.~T. Moynihan.
\newblock Description and analysis of electrical relaxation data for ionically
  conducting glasses and melts.
\newblock {\em Solid State Ionics}, 105:175--183, 1998.

\bibitem{DPIS97}
P.~Pissis and A.~Kyritsis.
\newblock Electrical conductivity in hydrogels.
\newblock {\em Solid State Ionics}, 97:105--113, 1997.

\bibitem{DMAC02}
J.~R. Macdonald.
\newblock Resolution of conflicting views concerning frequency--response models
  for conducting materials with dispersive relaxation, and isomorphism of
  macroscopic and microscopic models.
\newblock {\em Solid State Ionics}, 150:263--279, 2002.

\bibitem{hodge05}
I.~M. Hodge, K.~L. Ngai, and C.~T. Moynihan.
\newblock Comments on the electric modulus function.
\newblock {\em J. Non--Cryst. Solids}, 351:104--115, 2005.

\bibitem{DMUD00}
M.~Mudarra, J.~Belana, J.~C. Ca{\~n}adas, J.~A. Diego, J.~Sellar\`{e}s,
  R.~D\'{\i}az-Calleja, and M.~J. Sanchis.
\newblock Space charge relaxation in polyetherimides by the electric modulus
  formalism.
\newblock {\em J. Appl. Phys.}, 88:4807--4812, 2000.

\bibitem{lu06}
Hongbo Lu, Xingyuan Zhang, and Hui Zhang.
\newblock Influence of the relaxation of {Maxwell--Wagner--Sillars}
  polarization and {DC} conductivity on the dielectric behaviors of nylon 1010.
\newblock {\em J. Appl. Phys.}, 100:054104(7pp), 2006.

\bibitem{lanfredi02}
S.~Lanfredi, P.~S. Saia, R.~Lebullenger, and A.~C. Hernandes.
\newblock Electric conductivity and relaxation in fluoride, fluorophosphate and
  phosphate glasses: analysis by impedance spectroscopy.
\newblock {\em Solid State Ionics}, 146:329--339, 2002.

\bibitem{rivera01}
A.~Rivera, J.~Santamar\'{\i}a, and C.~Le{\'o}n.
\newblock Electrical conductivity relaxation in thin--film yttria--stabilized
  zirconia.
\newblock {\em Appl. Phys. Lett.}, 78:610--612, 2001.

\bibitem{liu03}
Jianjun Liu, Chun-Gang Duan, Wei-Guo Yin, W.~N. Mei, R.~W. Smith, and J.~R.
  Hardy.
\newblock Dielectric permittivity and electric modulus in
  {Bi}$_2${Ti}$_4${O}$_{11}$.
\newblock {\em J. Chem. Phys.}, 119:2812--2819, 2003.

\bibitem{psarras03}
G.~C. Psarras, E.~Manolakaki, and G.~M. Tsangaris.
\newblock Dielectric dispersion and {AC} conductivity in ---iron particles
  loaded--- polymer composites.
\newblock {\em Compos. Part A--Appl. S.}, 34:1187--1198, 2003.

\bibitem{migahed04}
M.~D. Migahed, M.~Ishra, T.~Fahmy, and A.~Barakat.
\newblock Electric modulus and {AC} conductivity studies in conducting {PP}y
  composite films at low temperature.
\newblock {\em J. Phys. Chem. Solids}, 65:1121--1125, 2004.

\bibitem{JONSCHER}
A.~K. Jonscher.
\newblock {\em Dielectric relaxation in solids}, chapter~5, pages 161--253.
\newblock Chelsea Dielectric Press, 1983.

\bibitem{JONSCHER2}
A.~K. Jonscher.
\newblock {\em Dielectric relaxation in solids}, chapter~8, pages 310--370.
\newblock Chelsea Dielectric Press, 1983.

\bibitem{DSID95B}
D.~L. Sidebottom, P.~F. Green, and R.~K. Brow.
\newblock Two contributions to the {AC} conductivity of alkali oxide glasses.
\newblock {\em Phys. Rev. Lett.}, 74:5068--5071, 1995.

\bibitem{DSID97}
D.~L. Sidebottom, P.~F. Green, and R.~K. Brow.
\newblock Scaling parallels in the non-debye dielectric relaxation of ionic
  glasses and dipols supercooled liquids.
\newblock {\em Phys. Rev. B}, 56:170--177, 1997.

\bibitem{LEO98}
C.~Le{\'o}n, M.~L. Luc\'{\i}a, J.~Santamar\'{\i}a, and F.~S\'anchez-Quesada.
\newblock Universal scaling of the conductivity relaxation in crystalline ionic
  conductors.
\newblock {\em Phys. Rev. B}, 57~(1):41--44, 1998.

\bibitem{anantha05}
P.~S. Anantha and K.~Hariharan.
\newblock {AC} conductivity analysis and dielectric relaxation behaviour of
  {NaNO}$_3$--{Al}$_2${O}$_3$ composites.
\newblock {\em Mat. Sci. Eng. B--Solid}, 121:12--19, 2005.

\bibitem{JONN80}
H.~M. Millany and A.~K. Jonscher.
\newblock Dielectric properties of stearic acid multilayers.
\newblock {\em Thin Solid Films}, 11:257--273, 1980.

\bibitem{DALM82}
D.~P. Almond, A.~R. West, and R.~J. Grant.
\newblock Temperature dependence of the {AC} conductivity of
  {N}a$\beta$--alumina.
\newblock {\em Solid State Commun.}, 44:1277--1280, 1982.

\bibitem{AVAK92}
H.~W.~Starkweather Jr. and P.~Avakian.
\newblock Conductivity and the electric modulus in polymers.
\newblock {\em J. Polym. Sci. Pol. Phys.}, 30:637--641, 1992.

\bibitem{DMAC}
J.~R. Macdonald.
\newblock {\em Impedance Spectroscopy: Emphasizing Solid Materials and
  Systems}, chapter~2, pages 27--132.
\newblock Wiley--Interscience, 1987.

\bibitem{LEO97}
C.~Le{\'o}n, M.~L. Luc\'{\i}a, and J.~Santamar\'{\i}a.
\newblock Correlated ion hopping in single--crystal yttria--stabilized
  zirconia.
\newblock {\em Phys. Rev. B}, 55~(2):882--887, 1997.

\bibitem{DLAR99}
E.~Laredo, N.~Suarez, A.~Bello, and J.~M.~G.~Fatou B.~R.~de G{\'{a}}scue,
  M.~A.~Gomez.
\newblock $\alpha$, $\beta$ and $\gamma$ relaxations of functionalized {HD}
  polyethylene: a {TSDC} and a mechanical study.
\newblock {\em Polymer}, 40:6405--6416, 1999.

\bibitem{DGRI97}
M.~Grimau, E.~Laredo, A.~Bello, and N.~Suarez.
\newblock Correlation between dipolar {TSDC} and {AC} dielectric spectroscopy
  at the {PVDF} glass transition.
\newblock {\em J. Polym. Sci. Pol. Phys.}, 35:2483--2493, 1997.

\bibitem{DBEL99}
A.~Bello, E.~Laredo, and M.~Grimau.
\newblock Distribution of relaxation times from dielectric spectroscopy using
  monte carlo simulated annealing: Application to $\alpha$-{PVDF}.
\newblock {\em Phys. Rev. B}, 60:12764--12774, 1999.

\bibitem{DSID95}
D.~L. Sidebottom, P.~F. Green, and R.~K. Brow.
\newblock Comparison of {KWW} and power law analyses of an ion--conducting
  glass.
\newblock {\em J. Non--Cryst. Solids}, 183:151--160, 1995.

\end{thebibliography}
\end{document}